\documentclass[pra, twocolumn]{revtex4}

\newcommand{\abs} [1] {\vert#1\vert} 
\newcommand{\e} [1] {\exp\sog{#1}} 
\newcommand{\half} {\frac{1}{2}} 
\newcommand{\imp} {\longrightarrow} 
\newcommand{\ox} {\otimes}
\newcommand{\prob} [1] {\mathrm{prob}\sog{#1}} 
\newcommand{\rhod} {\rho\mathrm{^{density}}} 
\newcommand{\scal} [2] {\langle#1\vert#2\rangle} 
\newcommand{\sog} [1] {\left(#1\right)} 
\newcommand{\tr} {\mathrm{tr}} 

\newcommand{\ket} [1] {\vert#1\rangle} 
\newcommand{\bra} [1] {\langle#1\vert} 
\newcommand{\uk} {\ket{\uparrow}} 
\newcommand{\ub} {\bra{\uparrow}} 
\newcommand{\dk} {\ket{\downarrow}} 
\newcommand{\db} {\bra{\downarrow}} 
\newcommand{\ak} [1] {\ket{\mathrm{#1}}} 
\newcommand{\ab} [1] {\bra{\mathrm{#1}}} 
\newcommand{\ek} [1] {\ket{\mathrm{\varepsilon_{#1}}}} 
\newcommand{\eb} [1] {\bra{\mathrm{\varepsilon_{#1}}}} 

\begin{document}

\title{Two-time interpretation of quantum mechanics}
\author{Yakir Aharonov}
\author{Eyal Y. Gruss}\email[e-mail: ]{eyalgruss@gmail.com}
\affiliation{School of Physics and Astronomy,\\Beverly and Raymond
Sackler Faculty of Exact Sciences,\\Tel-Aviv University, Tel-Aviv
69978, Israel.}

\begin{abstract}
We suggest an interpretation of quantum mechanics, inspired by the
ideas of Aharonov et al. of a time-symmetric description of
quantum theory. We show that a special final boundary condition
for the Universe, may be consistently defined as to determine
single classical-like measurement outcomes, thus solving the
``measurement problem''. No other deviation is made from standard
quantum mechanics, and the resulting theory is deterministic (in a
two-time sense) and local. Quantum mechanical probabilities are
recovered in general, but are eliminated from the description of
any single measurement. We call this the \emph{Two-time
interpretation} of quantum mechanics. We analyze ideal
measurements, showing how the quantum superposition is, in effect,
dynamically reduced to a single classical state via a ``two-time
decoherence'' process. We discuss some philosophical aspects of
the suggested interpretation. We also discuss weak measurements
using the two-time formalism, and remark that in these measurement
situations, special final boundary conditions for the Universe,
might explain some unaccounted for phenomena.
\end{abstract}

\maketitle

\section{Introduction}
The ``measurement problem'' is a most fundamental problem in the
field of the foundations of physics. The problem is that while
quantum mechanics allows very accurate calculation of microscopic
phenomena, it is not clear how it can fully describe macroscopic
measurement processes. Empirically, a measurement performed on a
quantum system yields one single outcome, with some probability
given by quantum mechanics. By contrast, in the standard unitary
quantum mechanics formalism, the quantum state evolves into a
linear superposition of the possible measurement outcomes. This
discrepancy is a bothersome loophole in the theory, and it can be
an actual difficulty when considering ``theories of everything''
based on quantum mechanics.

To bridge the gap between theory and observed reality, different
interpretations of quantum mechanics, or rather, different
interpretations of our observations, have been suggested. Some of
these interpretations abandon conventional classical concepts such
as determinism and locality, which we briefly define.

``Locality'', means that an event cannot have any influence
outside its future light-cone (even if such influence does not
defy relativistic causality, i.e., does not allow superluminal
signalling, it is still forbidden). Since nonlocality defies
relativistic covariance, it is usually considered an undesirable
property of a theory.

``Determinism'' means that the physical state at any time is
determined completely as a single-valued evolution of the state at
some single (usually initial) time. The validity of this principle
constitutes a main disagreement point between the different
interpretations of quantum mechanics, and therefore we shall
classify them by this criterion.

Usually, indeterministic interpretations involve a ``collapse''
phenomenon. The quantum mechanics formalism is modified by the
addition of a wave function collapse rule, responsible for the
reduction of the superposition into a single state. This approach
has several weaknesses: First, indeterminism is of course
counterintuitive to the conventional deterministic perception of
nature. Second, the interpretation is nonlocal, as the collapse is
assumed to take place instantaneously in all space. This is
clearly a noncovariant description, since collapse events that are
simultaneous in some frame of reference would not be simultaneous
in any boosted frame of reference. Ref. \cite{noncovar} discusses
some of the dilemmas of formulating a consistent covariant
collapse scheme. Third, most collapse theories lack well defined
mechanism and criteria for the collapse process. Such is the case
with the orthodox collapse approach, which states merely that a
collapse occurs ``somewhere'' along the line of the measurement
process. Nevertheless, more rigorous suggestions do exist, such as
the GRW spontaneous localization model and successors thereof
\cite{grw}.

Some deterministic interpretations are based on Bohmian ``hidden
variables'' \cite{bohm}. These assume the existence of
inaccessible local variables with definite values which determine
the measurement outcome. Bell's celebrated inequality \cite{bell},
and the more recent GHZ argument \cite{ghz}, show that hidden
variable theories are inherently nonlocal. A different
deterministic interpretation is Everett's ``relative state''
formulation \cite{everett}, also known as the ``many worlds''
interpretation. Here the state after the measurement is considered
to be the full superposition state, where it is assumed that all
measurement outcomes in the superposition coexist as separate real
world outcomes. The observation of a certain outcome is attributed
to the specific state of the observer that is correlated to it in
the superposition. Each of the superposition terms constitutes a
``branching world'', whereas the overall state evolves unitarilly
and is deterministic. In this interpretation it is not clear how
to recover the empirical quantum mechanical probabilities. Both of
the above deterministic interpretations require additional
entities (wave function plus hidden variables, many worlds) in
order to achieve the complete description of physical nature.
These entities might be considered an excess by Occam's razor, the
simplicity postulate, which states that ``Entities are not to be
multiplied beyond necessity''.

We do not consider here epistemological interpretations (including
the ``Copenhagen'' interpretation and some ``consistent
histories'' interpretations), as opposed to the ontological
interpretations above. The former pretend only to give a set of
logical rules regarding our knowledge of reality, and do not
attempt to explain any underlying processes taking place.

In this work we suggest an ontological interpretation, which
attempts to overcome the difficulties mentioned above. It is
deterministic, local, and simple, and it recovers the empirical
probabilities of quantum mechanics.

To complete the required background, we briefly discuss the
notions of ``classicality'' and ``decoherence''. After the
measurement interaction between the quantum system and the
measuring device (of course also a ``quantum'' system) is
complete, and assuming no reduction takes place, we get a
superposition of states of possible outcomes. By a ``state of a
possible outcome'' we mean a measuring device state correlated to
the corresponding system state. Our experience tells us, these
measuring device states belong to a certain definite basis of
localized states. No outcome in the form of a superposition of
these states can be measured. We regard these states as
``classical'' states and postulate that it is impossible to
interfere or ``mix'' these states. This is usually assumed to be
so due to a decoherence process a la Zurek \cite{zurek}. In this
process the essential part of the measuring device, referred to as
the ``pointer'', becomes correlated to additional quantum systems,
the ``environment''. It is assumed that the environment states,
due to their interaction with the pointer states, very quickly
become and remain nearly orthogonal. The correlation of the
pointer with these environment states defines a preferred basis of
pointer states, in relation to the original measurement. Further,
it is assumed that the environment degrees of freedom are
practically inaccessible, and thus they must be ignored. Tracing
out these degrees of freedom, we obtain the reduced density
matrix, which shows to have negligible off-diagonal interference
terms. We thus get a dynamical superselection process, which damps
out any superpositions of classical states, isolating the
classical states in an effectively irreversible way. In this work,
we characterize ideal measurements as those in which the
measurement outcome (or a record thereof) remains classical up to
some very far ``final time''. Of course, decoherence by itself
does not resolve the measurement problem -- no reduction to a
single classical state occurs. Yet, most importantly, a preferred
basis for such a reduction is defined and superselected.

The outline of the article is as follows: We start with a brief
review of the two-time, or time-symmetric, formalism of quantum
mechanics. We suggest a generalization of this formalism in order
to describe the measurement process in closed systems. This is
done by the introduction of a special final boundary condition for
the Universe, which accounts for observed measurement outcomes. We
analyze ideal measurements in the framework of the suggested
interpretation, and show how effective reduction arises
dynamically due to the final boundary condition, with no
additional mechanism. We call this process ``two-time
decoherence''. Next, we briefly discuss some philosophical aspects
of the suggested interpretation. We address the relevance and
validity of concepts such as locality, microscopic
irreversibility, causality, determinism, freedom of choice,
realism and counterfactual definiteness, in the framework of our
interpretation. Finally, we discuss the concept of non-ideal
``weak measurements''. The outcome of which is a two-time
expectation value of the measured operator, the ``weak value''. We
explain the emergence of the weak value using the two-time
formalism, and discuss the possible effect of a special final
boundary condition for the Universe, in this context.

\section{Generalization of the Two-Time Formalism to Closed Systems}
In 1964, ABL \cite{abl} derived a probability rule concerning
measurements performed on systems, with a final state specified in
addition to the usual initial state. Such a final state may arise
due to a post-selection, that is, performing an additional
measurement on the system and considering only the cases with the
desired outcome. Alternatively, some systems in nature may have an
inherent final boundary condition, just as all systems have an
initial boundary condition. Given an initial state $\Psi_i$ and a
final state $\Psi_f$, the probability that an intermediate
measurement of the non-degenerate operator $A$ yields an
eigenstate $a_k$ is
\begin{widetext}
\begin{equation}
\label{eq:abl} \prob{a_k\mid\Psi_i,\Psi_f}= \frac{\prob{\Psi_f\mid
a_k}\prob{a_k\mid\Psi_i}} {\sum_j{\prob{\Psi_f\mid
a_j}\prob{a_j\mid\Psi_i}}}=
\frac{\abs{\scal{\Psi_f}{a_k}}^2\abs{\scal{a_k}{\Psi_i}}^2}
{\sum_j\abs{\scal{\Psi_f}{a_j}}^2\abs{\scal{a_j}{\Psi_i}}^2}.
\end{equation}
\end{widetext}
For simplicity, no self-evolution of the states is considered
between the measurements. If only an initial state is specified,
(\ref{eq:abl}) should formally reduce to the regular probability
rule:
\begin{equation}
\label{eq:regprob}
\prob{a_k\mid\Psi_i}=\abs{\scal{a_k}{\Psi_i}}^2.
\end{equation}
This is usually shown by summing over the probabilities of all
possible final states, expressing the indifference to the final
state. However, (\ref{eq:regprob}) can also be obtained in a
different way. If the final state would be one of the eigenstates,
$\Psi_f=a_k$, then ABL (\ref{eq:abl}) gives probability one for
measuring $a_k$, and probability zero for measuring any orthogonal
state. Consider now an ensemble of systems of whom fractions of
size $\prob{a_k\mid\Psi_i}$ happen to have the corresponding final
states $a_k$. The regular probability rule (\ref{eq:regprob}) of
quantum mechanics is then recovered. But now the probabilities are
classical probabilities due to ignorance of the specific final
states, and the description is deterministic in a two-time sense.
The same would be the result for a corresponding final state of an
auxiliary system, such as a measuring device or environment,
correlated with the measured system. This reduction of the ABL
rule to the regular probability rule is a clue, showing how a
selection of appropriate final states can account for the
empirical probabilities of quantum measurements.

It is clear from the ABL rule that the final state of a system may
be of importance. Weak measurements, which are discussed later,
are one example. Following \cite{twostate}, we reformulate quantum
mechanics to be time-symmetric, in the sense that it will take
into account both initial and final boundary conditions. The
Schr\"odinger equation is linear in the time derivative, therefore
only one temporal boundary condition may be consistently specified
for the wave function. If both initial and final boundary
conditions exist, we must have two wave functions, one for each
boundary condition. The first is the regular wave function, or
state vector, evolving forward in time from the initial boundary
condition. We call it the ``history vector'',
$\Psi_\mathrm{HIS}(t)$. The second is a different wave function
evolving from the final boundary condition backwards in time,
which we call the ``destiny vector'', $\Psi_\mathrm{DES}(t)$. A
measurement (including post-selection) will later be shown to
constitute an effective boundary condition for both wave
functions. We postulate that the complete description of a closed
system is given by the two wave functions. These may be combined
into operator form by defining the ``two-state'':
\begin{equation}
\rho(t)\equiv\frac{\ket{\Psi_\mathrm{HIS}(t)}\bra{\Psi_\mathrm{DES}(t)}}%
{\scal{\Psi_\mathrm{DES}(t)}{\Psi_\mathrm{HIS}(t)}},
\end{equation}
where orthogonal history and destiny vectors at any time $t$ are
forbidden. For a given Hamiltonian $H(t)$, the two-state evolves
from time $t_1$ to $t_2$ according to
\begin{equation}
\rho(t_2)=U(t_2,t_1)\rho(t_1)U(t_1,t_2),
\end{equation}
where $U(t_2,t_1)$ is the regular evolution operator:
\begin{equation}
U(t_2,t_1)=T\e{-i/\hbar\int_{t_1}^{t_2}H(\tau)d\tau}
\end{equation}
($T$ signifies the time ordered expansion). The reduced two-state
describing a subsystem, is obtained by tracing out the irrelevant
degrees of freedom.

In standard quantum mechanics we may also use operator form
similar to the above, replacing the state vector
$\Psi_\mathrm{HIS}(t)$ with the density matrix:
\begin{equation}
\rhod(t)\equiv\ket{\Psi_\mathrm{HIS}(t)}\bra{\Psi_\mathrm{HIS}(t)}.
\end{equation}
The density matrix again evolves by
\begin{equation}
\rhod(t_2)=U(t_2,t_1)\rhod(t_1)U(t_1,t_2),
\end{equation}
and again the reduced density matrix for a subsystem, is obtained
by tracing out the irrelevant degrees of freedom. Assuming unitary
evolution (no measurements), the density matrix is a complete
description of systems which, as is usually assumed, evolve from
some initial state and (apparently) have no definite final
boundary condition. For such systems, the two-time formalism
should reduce to the standard one. This will be so, if the destiny
vector is equal to the history vector at any time,
\begin{equation}
\ket{\Psi_\mathrm{DES}(t)}=\ket{\Psi_\mathrm{HIS}(t)}.
\end{equation}
This is true for any system with a trivial final boundary
condition that is just the initial state evolved unitarilly from
the initial time $t_i$ to the final time $t_f$,
\begin{equation}
\label{eq:f=ui}
\ket{\Psi_f}=U(t_f,t_i)\ket{\Psi_i}.
\end{equation}
Therefore, apart from the measurement problem, it is clear that
the standard formalism is a special case of the two-time
formalism. We trivially take (\ref{eq:f=ui}) as a zero order
approximation of the final boundary condition. The ABL rule
(\ref{eq:abl}) is not relevant here (it would give a square of the
regular probability), since we are not yet considering
measurements. Now by considering final boundary conditions
deviating from the above, we may introduce a richer state
structure into quantum theory. When would the final boundary
condition and two-time formalism show to affect the dynamics? It
would do so if the reduced two-state describes a subsystem for
which the ignored degrees of freedom do not satisfy
(\ref{eq:f=ui}). Then, the reduced two-state should replace the
density matrix which is no longer a reliable description of the
state of the system.

A measurement, as empirically observed, generally yields a new
outcome state of the quantum system and the measuring device. This
state may be treated as an effective boundary condition for both
future, and as indicated by the ABL rule, past events. We suggest
that it is not the case that a new boundary condition is generated
at each measurement event, by some unclear mechanism. Rather, only
a final boundary condition needs to be given for the measuring
device, as part of a final boundary condition of the Universe. In
the following section we shall demonstrate how an effective
boundary condition then arises at the time of measurement due to a
two-time decoherence effect. ABL (\ref{eq:abl}) of course agrees
that in the classical basis (determined by decoherence), the
outcome of the measurement can only be the single classical state
corresponding to the final boundary condition.

We thus suggest a special final boundary condition for the
Universe, in which each classical system (measuring device) has a
final boundary condition equal to one of its possible classical
states (evolved to the final time). We further assume that these
final states have an appropriate distribution so as to recover the
empirical quantum mechanical probabilities for large ensembles.
The final boundary condition is a boundary condition for the
destiny vector (not the history vector). Since the description is
unitary, the destiny vector data may be given at any time, and we
may ignore the question of the actual cosmological final state of
the Universe. A scheme to choose the final boundary condition
would be as follows: Given the initial boundary condition of the
Universe and the Hamiltonian, calculate the trivial final boundary
condition as the initial boundary condition evolved unitarilly to
the final time. Next, identify the classical systems and their
classical basis, due to the effect of decoherence. Change the
trivial final boundary condition, so that the final boundary
condition of each classical system (and systems correlated with
it) is a single normalized term of the calculated final
superposition state written in the classical basis. The choice of
the specific state should be random, with a probability
proportional to the squared amplitude of the corresponding term in
the superposition. With this choice, the two-time formalism of
\cite{twostate} is generalized to fully describe the measurement
process in closed systems, and there is no longer a ``measurement
problem''. The measurement outcome is selected, as we shall show,
by the dynamics due to unitary Schr\"odinger evolution alone. Note
that the measurement outcome will be realized in a subsystem which
is ``open'', but a complete description may be given for any
closed system, namely for the closed Universe.

After the completion of this work, we found a paper by Davidon
\cite{davidon} dated 1976, which suggests a description similar to
ours.

\section{Ideal Measurements}
Consider an experiment performed on a spin-$\half$ particle, in
order to measure its spin component along some axis. Let the
initial state of the particle be $\sog{a\uparrow+b\downarrow}$,
and denote the initial state of the measuring device $READY$, and
the initial state of the environment $\varepsilon_{0}$. The
initial state, or the history vector, of the composite system at
the initial time $t_0$ is
\begin{equation}
\ket{\Psi_\mathrm{HIS}(t_0)}=\sog{a\uk+b\dk}\ox\ak{READY}\ox\ek{0}.
\end{equation}
Assume an interaction between the particle and the measuring
device takes place until time $t_1$, such that if the device
measures $\uparrow$, it evolves into the state $UP$, and if it
measures $\downarrow$, it evolves into the state $DOWN$. The
composite system then evolves to the state:
\begin{equation}
\ket{\Psi_\mathrm{HIS}(t_1)}=\sog{a\uk\ox\ak{UP}+b\dk\ox\ak{DOWN}}\ox\ek{0}.
\end{equation}
Assume that after this time, a decoherence process takes place, in
which the measuring device interacts with the environment, giving
after a short decoherence time $t_d$,
\begin{eqnarray}
\ket{\Psi_\mathrm{HIS}(t>t_1+t_d)}&=&a\uk\ox\ak{UP}\ox\ek{up}+\nonumber\\*
&+&b\dk\ox\ak{DOWN}\ox\ek{down},\nonumber\\*
\end{eqnarray}
where $\varepsilon_{up}$ and $\varepsilon_{down}$ are nearly
orthogonal environment states, which induce the superselection of
$UP$ and $DOWN$ as the preferred basis of pointer states. By our
definition of an ideal measurement, decoherence is assumed to
cause these states to remain classical up to some far ``final
time''. For the time being, assume that after the measurement
interaction is over, the measuring device is left idle and its
state remains unchanged. We now introduce the novel key element of
the suggested interpretation: We assume a specific final boundary
condition for the measuring device and correlated environment at
the above final time. Let this be
\begin{equation}
\bra{...}\ox\ab{UP}\ox\eb{up},
\end{equation}
where ``...'' represents the state of any other system correlated
with the measuring device at the final time. The final boundary
condition was chosen as one specific state of the preferred basis.
This is legitimate since decoherence prevents any interference
with this classical state up to the final time. The particle, by
contrast, remains in its measured state only until some other
measurement is performed on it, preparing it in a new arbitrary
state:
\begin{equation}
\ket{\phi}=c\uk+d\dk.
\end{equation}
For the time being, assume that this second measurement takes
place instantaneously at time $t_2$, producing the
backward-evolving state $\phi$. This takes the role of a final
boundary condition for our particle. We can write the effective
final boundary condition, or the destiny vector, of the composite
system at time $t_2$ as
\begin{equation}
\bra{\Psi_\mathrm{DES}(t_2)}=\bra{\phi}\ox\ab{UP}\ox\eb{up}.
\end{equation}
At any time $t$: $t_1+t_d<t<t_2$ the complete description of the
composite system is given by the two-state (up to normalization):
{\setlength\arraycolsep{0pt}
\begin{eqnarray}
\rho(t)&=&\ket{\Psi_\mathrm{HIS}(t)}\bra{\Psi_\mathrm{DES}(t)}=\nonumber\\*
&=&a\uk\ox\ak{UP}\ox\ek{up}\bra{\phi}\ox\ab{UP}\ox\eb{up}+\nonumber\\*
&+&b\dk\ox\ak{DOWN}\ox\ek{down}\bra{\phi}\ox\ab{UP}\ox\eb{up}.\nonumber\\*
\end{eqnarray}}
Tracing out the environment degrees of freedom, we obtain the
reduced two-state describing the particle-measuring device
subsystem:
\begin{equation}
\rho_{reduced}(t)=\tr_\varepsilon\rho(t)\approx
\uk\bra{\phi}\ox\ak{UP}\ab{UP}.
\end{equation}
One may say that an effective reduction has occurred, yielding the
single measurement outcome: ``UP''. The reduction is the result of
a two-time environment induced superselection process. We call
this process, which dictates the classical behavior of the
measuring device, ``two-time decoherence'', in analogy with the
regular notion.

We have shown how effective reduction can take place subsequent to
the measurement process. The process is actually time-symmetric.
Remember we have treated the measurement following our measurement
(the one that yielded $\phi$) as instantaneous, in order to
formulate the backward-evolving state. We now relax this
assumption by showing how effective reduction occurs for, and
determines, the backward-evolving state of the particle. Evolving
the destiny vector backward in time to $t_1$, we obtain:
\begin{equation}
\bra{\Psi_\mathrm{DES}(t_1)}=\bra{\phi}\ox\ab{UP}\ox\eb{0},
\end{equation}
and at $t_0$:
\begin{eqnarray}
\bra{\Psi_\mathrm{DES}(t_0)}&=&\sog{c\ub\ox\ab{READY}+d\db\ox\ab{ORTHO}}\ox\nonumber\\*
&\ox&\eb{0},
\end{eqnarray}
where the time-reversed interaction between the measuring device
and the particle, causes a device in the final state $UP$ to
evolve backwards into the state $READY$, if the particle is in the
state $\uparrow$, and into the orthogonal state $ORTHO$, if the
particle is in the state $\downarrow$. Again we assume that an
environment induced decoherence process takes place (here
backwards in time, but the microscopic physics is time-symmetric),
singling out $READY$ and $ORTHO$ as the preferred basis of pointer
states for the destiny vector. Then the reduced two-state at times
$t<t_0-t_d$ (where $t_d$ is the decoherence time) is
\begin{eqnarray}
\rho_{reduced}(t<t_0-t_d)&=&\sog{a\uk+b\dk}\ub\ox\nonumber\\*
&\ox&\ak{READY}\ab{READY}.
\end{eqnarray}
Now the backward-evolving state of the particle is $\uparrow$, as
is expected to evolve backwards in time from our ``UP''
measurement. This sets a final boundary condition for any previous
measurement performed on the particle. Since the information for
the reduction of the backward-evolving state is carried by the
measuring device destiny vector, we see that a final boundary
condition is required for the measuring device itself. A final
boundary condition for the environment alone, as one might
suggest, would be insufficient. The forward-evolving state of the
particle before the time of the measurement, is of course not
affected by the final boundary condition.

When considering backward-in-time evolution, one may be concerned
about issues of stability. Refer to the environment states
$\varepsilon_{up}$ and $\varepsilon_{down}$ used above, at the
decoherence time $t_1+t_d$ (which will serve as the initial time
for the current example). Assume these states consist of clusters
of $N$ particles, concentrated respectively in the upper and
bottom tenth of an isolated one-dimensional box of one meter
length. The particles of each macrostate evolve as to scatter and
spread out in the box, the two macrostates remaining nearly
orthogonal to each other. As described before, a final boundary
condition is assumed in the form of a measuring device pointer
state coupled to, and superselected by, one of these environment
macrostates, say $\varepsilon_{up}$ (all states in their form at
the final time). This destiny vector evolves backwards in time to
determine the outcome in the measuring device. Now, let the system
not be completely isolated, and allow external quantum
disturbances which interfere with the evolution in a causally
indeterministic and thus irreversible way. While the naked eye
might not notice the effect on the late time spread out
macrostates, a random disturbance of the time-reversed evolution
would obviously give rise to initial states very different from
the original concentrated states. This is the instability of the
time-reversed evolution, a process of decreasing entropy due to
the second law of thermodynamics. The question then rises, what
are the implications on the suggested interpretation, where the
backward-evolving state determines the measurement outcome. The
threat will come from state changing quantum measurements
performed on the environment particles. Consider therefore, a
position measurement performed on $n$ of the particles, with a
precision of one angstrom, at a time when the particles are spread
out in the box. The new disturbed macrostates: $\varepsilon'_{up}$
and $\varepsilon'_{down}$, have projection amplitudes:
$\scal{\varepsilon'_{up\;(down)}}{\varepsilon_{up\;(down)}}=10^{-10n}$.
Assuming our environment is large enough in order to constitute a
good environment, the disturbed macrostates are still nearly
orthogonal, and the decoherence process remains intact. We now
have to consider the implication of having $\varepsilon'_{up}$
instead of $\varepsilon_{up}$ in the destiny vector, evolving
backwards from the disturbance measurement. The above requirement
for the intactness of decoherence, implies that the disturbed
state is still nearly orthogonal to $\varepsilon_{down}$. On the
other hand, the scalar product with $\varepsilon_{up}$, on which
the disturbed state was projected, is of course finite
($10^{-10n}$). For a crude upper limit on the scalar product with
$\varepsilon_{down}$, one may take $10^{-N}$, accounting for the
spread out state of the $\varepsilon'_{up}$ at the initial time,
at which $\varepsilon_{down}$ is concentrated at the bottom tenth
of the box. In order for the two-time decoherence scheme to still
work, the latter number should be negligible relative to the
former, that is $N\gg10n$. This is a requirement on the size of
the environment relative to the disturbance, which is presumably
connected to the requirement for decoherence.

Our example can readily be extended to more complex systems. For
instance, we have considered only a single measurement process per
measuring device, where it is of course possible to perform more
than one measurement with the same device. Since the measuring
device has a single specific final state (assuming it has only one
degree of freedom), where does the information for the many
measurement outcomes reside? The answer is that a device that is
used for multiple measurements, must be initialized between the
measurements. The unitary initialization interaction transfers the
information of the previous state of the measuring device to other
systems, no information being lost. In any case, a multiple-time
generalization of the two-time description, with a multitude of
intermediate boundary conditions (such as the picture in
``consistent histories'' interpretations) is not required in order
to account for multiple measurements.

\section{Philosophical Aspects}
It was already assumed, that subsequent to the measurement
interaction, decoherence causes an effectively irreversible
branching of the superposition into isolated terms. Therefore, no
inconsistencies can arise from the existence of a special final
boundary condition of the form described before, which simply
causes the selection of a single specific branch from the many
worlds picture. For this reason, the locality property of standard
quantum mechanics remains valid in the suggested interpretation.

Also for the above reason, the measurement process does not
increase the measure of irreversibility beyond that of regular
thermodynamics. That is, the suggested interpretation does not
suggest a microscopic quantum mechanical arrow of time. It does
however assume asymmetric initial and final boundary conditions.

It must be emphasized that the states in the final boundary
condition, which we have taken to be specific in the examples, are
generally unknown prior to the completion of the measurement.
Classically, a priori knowledge of the future is of course an
acausal state of affairs. The following example shows that it is
also a problem due to quantum mechanics itself. Assume there are
two entangled spin-$\half$ particles located at two far away
locations, in the initial state
$\uparrow_A\uparrow_B+\downarrow_A\downarrow_B$. $A$ denotes the
particle at Alice's location, and $B$ -- the particle at Bob's
location. Assume that Bob knows the final state to be
$\uparrow_A\uparrow_B+\uparrow_A\downarrow_B$. Now, Alice may or
may not perform a unitary rotation on her particle, of the form
$\uparrow_A\imp\downarrow_A$ and $\downarrow_A\imp\uparrow_A$,
leaving the initial composite state as it was, or transforming it
into the state $\downarrow_A\uparrow_B+\uparrow_A\downarrow_B$.
Bob, measures the spin of his particle, obtaining $\downarrow_B$
or $\uparrow_B$, according to the action or non-action of Alice.
In this manner, Alice may allegedly transmit signals to Bob at an
instant. A procedure like this would be possible for many
arbitrary choices of initial and final states. Only when identical
measurements are performed in sequence, can a final state (or a
measurement outcome) be predicted with certainty in advance.

Therefore, like in hidden variables theories, the parameters
determining the measurement outcome are inaccessible. Still, like
in those theories, the evolution may be considered deterministic
(though unpredictable). As mentioned before, determinism is valid
if considered in a broader two-time context, where the evolution
is determined not only by an initial boundary condition, but also
by a final boundary condition. The latter dynamically determines
the measurement outcomes, such that in all intermediate times the
physical states are determined by (two) unitary Schr\"odinger
evolution. Given the boundary conditions and a Hamiltonian, one
may reconstruct the whole evolution history, no random dice need
be tossed. Alternatively, the state of the system at any time, is
completely determined by its two-state at any single time.

The existence of a future boundary condition, and its
deterministic effect, do not deny our apparent freedom of choice.
The latter is allowed due to the inaccessibility of the data
(which was a requirement of causality). Imagine an external
observer with a reversed arrow of time, classically (or weakly)
monitoring and recording our future measurements, without
disturbing them. As long as he does not pass to us any beforehand
information, the picture is consistent by definition. His records,
analogous to a future boundary condition, cannot be spoiled by
(nor do they prohibit) a ``change of mind'' on our behalf, which
would already be taken into account. ``Foreknowledge no more
``forces'' the future to be a certain way, than true reports in
history books ``force'' the past to have been a certain
way''\cite{swartz}. Perhaps this is close to the omniscient
approach expressed by the old Hebrew sages: ``All is foreseen and
the choice is given''\cite{sages}.

So, our apparent freedom of choice is not at danger. But is a
genuine freedom of choice really possible? It is not easy to see
how such a concept could even be well defined in physical terms.
Nevertheless, consider the case that the extra degree of freedom
of an agent-system, the destiny vector, corresponds to what we
refer to as the ``free choice of the agent''. This determines the
outcomes of (cerebral) quantum measurements, in the manner
described in this work. We have a description where the choice is
``causa sui'' (cause of itself), while still being the choice of
the agent-system. This would constitute a unique realization of
the concept of genuine free will.

The last property we wish to address is that of ``realism'' or
``objectivity''. These refer to the classical concept that the
existence of physical properties is independent of observations of
these properties. EPR \cite{epr} define realism by the following
counterfactual:
\begin{quotation}
If, without in any way disturbing a system, we can predict with
certainty (i.e., with probability equal to unity) the value of a
physical quantity, then there exists an element of physical
reality corresponding to this physical quantity.
\end{quotation}
This would require the existence of some additional (possibly
``hidden'') variables, which determine the outcomes of the
measurements. As mentioned before, hidden variable theories are
inherently nonlocal \cite{bell},\cite{ghz}, and the possibility of
local realism is excluded. This line of reasoning requires the
validity of ``counterfactual definiteness''. The meaning of which
is that it is meaningful to ask hypothetic ``what-if'' questions.
If it is not, the EPR definition of realism is irrelevant. Such is
the case with the many worlds interpretation, where each
measurement yields all possible outcomes. This is also the case
with the suggested interpretation, were it is assumed that a final
boundary condition is predetermined according to the measurements
that actually take place. The destiny vector constitutes a special
element of reality or hidden variable, which answers only the
question that is being asked.

\section{Weak Measurements}
Aharonov et al. have introduced the concept of weak measurements
and weak values \cite{weakval}. The idea is to perform a
measurement, in which the interaction is weak enough, in some
sense, leaving the two-state (both history and destiny vectors) of
the measured system almost undisturbed. The weakness of the
interaction, implies that the state of the measuring device cannot
be sharp but rather has a broad spread (large uncertainty) in its
pointer variable. The weighted superposition of the shifted broad
pointer states, that are obtained by the interaction, adds up to a
single state centered about a two-time expectation value of the
measured operator. This is the weak value of the measured
operator, which is determined by both initial and final states of
the measured system. For a system with initial state $\Psi_i$ and
final state $\Psi_f$, the weak value of operator $A$ is given by
\begin{equation}
A_w\equiv\frac{\bra{\Psi_f}A\ket{\Psi_i}}{\scal{\Psi_f}{\Psi_i}}.
\end{equation}
In general $A_w$ may be far from any eigenvalue of $A$. The
measurement is concluded with an observation of the measuring
device, effectively reducing its state to give a sharp outcome
according to its probability distribution. Weak values may play an
important role in the interpretation of certain phenomena such as
tunnelling \cite{tunnelling}, trans-Planckian frequencies in the
Hawking radiation from a black hole \cite{hawking} and Cherenkov
radiation of superluminal particles \cite{cherenkov}. A formal
reduction to the single-time case is made by considering a final
state equal the initial state, $\psi_f=\psi_i$. This is just the
trivial boundary condition (\ref{eq:f=ui}), assuming that the
measured system is almost undisturbed by the measurement, and
assuming for simplicity no self-evolution during the measurement
process. The weak value then reduces to the regular expectation
value of the measured operator $A$,
\begin{equation}
\label{eq:expval} \langle
A\rangle\equiv\bra{\Psi_i}A\ket{\Psi_i}=\sum_ka_k\prob{a_k\mid\Psi_i}.
\end{equation}

Let us follow the weak measurement process using the two-time
formalism. Consider a many level quantum system in the basis
$\{a_k\}$, and the operator $A$ defined as
\begin{equation}
A=\sum_ka_k\ket{a_k}\bra{a_k}.
\end{equation}
We take the initial state of the system to be
\begin{equation}
\ket{\phi_1}=\sum_kc_k\ket{a_k}.
\end{equation}
The measuring device is described by a pointer state $Q(q)$, which
is a Gaussian-like function centered around the pointer variable
$q$. The initial state of the measuring device is $Q(0)$, and the
initial state of the composite system is
\begin{equation}
\ket{\Psi_i(0)}=\ket{\phi_1}\ox\ket{Q(0)}.
\end{equation}
A measurement of $A$ is performed by an interaction between the
system and the measuring device, such that if the device measures
$a_k$, it evolves into the state $Q(a_k)$. The state of the
composite system at time $t_1$ after the measurement interaction
is
\begin{equation}
\ket{\Psi_i(t_1)}=\sum_kc_k\ket{a_k}\ox\ket{Q(a_k)}.
\end{equation}
At a later time $t_2$ we perform an ideal post-selection
measurement on the quantum system obtaining the result:
\begin{equation}
\ket{\phi_2}=\sum_kc_k'\ket{a_k},
\end{equation}
which serves as a final boundary condition for the measured
system. The final composite state is thus the projection on
$\phi_2$:
\begin{equation}
\ket{\Psi_f(t_2)}=\ket{\phi_2}\bra{\phi_2}\ket{\Psi_i(t_1)}=%
\ket{\phi_2}\ox\sum_kc_kc_k'^*\ket{Q(a_k)}.
\end{equation}
The condition for the weakness of the measurement is that the
Gaussians in the last term are sufficiently broad so that the
relation
\begin{equation}
\sum_kc_kc_k'^*\ket{Q(a_k)}\approx\sum_kc_kc_k'^*\ket{Q(a')}\equiv\ket{\hat{Q}(a')},
\end{equation}
holds for some $a'$. It can be shown \cite{weakval} that $a'$ is
just $A_w$, the weak value of $A$,
\begin{equation}
A_w=\frac{\bra{\phi_1}A\ket{\phi_2}}{\scal{\phi_1}{\phi_2}}=%
\frac{\sum_kc_kc_k'^*a_k}{\sum_kc_kc_k'^*}.
\end{equation}
Therefore the measuring device reading is centered on the real
part of the weak value. It is clear that after post-selection, the
weak value is obtained as a consequence of the projection onto the
new quantum state $\phi_2$. Looking at the reduced two-state of
the measuring device, it is easy to see how the weak value
actually emerges before the post-selection measurement is
performed. The two-state of the composite system at a time $t$
between $t_1$ and $t_2$ (after the weak interaction is complete
and before post-selection) is given (up to normalization) by
\begin{equation}
\rho(t)=\ket{\Psi_i(t_1)}\bra{\Psi_f(t_2)}.
\end{equation}
Tracing out the measured system degree of freedom, the measuring
device already reads the weak value:
\begin{eqnarray}
\rho_{pointer}(t)=\tr_\phi\rho(t)&=&\sum_{k,j}c_kc_k'^*c_j^*c_j'\ket{Q(a_k)}\bra{Q(a_j)}\approx\nonumber\\*
&\approx&\ket{\hat{Q}(A_w)}\bra{\hat{Q}(A_w)}.
\end{eqnarray}

Considering a complete set of post-selections, the probability
distributions of the corresponding weak measuring device states,
add up to recover the regular probability distribution without
post-selection. The probability distribution of any specific state
of the former is always smaller than the latter, and the
post-selection is consistent with the standard quantum mechanical
probabilities (where the ``weak'' result may always be considered
a ``measurement error''). Also, the weakness condition must imply
that the initial state of the measuring device is, in general, an
analytic function in a strip around the real axis of the pointer
variable. Thus any local region contains the information of the
shape of the entire wave function even before the interaction
takes place, and the process is one of amplification rather than
information transfer. The above two properties of weak
measurements ensure that the procedure is always causally
consistent.

Measurements performed on very large macroscopic systems are
essentially weak measurements, which examine the system's past and
future boundary conditions. Consider, for example, measurements of
galactic properties such as mass or angular momentum. A consistent
``strange'' measurement outcome, might arise due to a special
final boundary condition of the measured system, different from
the trivial boundary condition (\ref{eq:f=ui}). This would result
in the observation of a weak value which may be far from the
expectation value. In contrast to the case of a post-selection
measurement, the effect of a special natural final boundary
condition would seem to us as new fundamental laws of nature, here
already breaking the framework of standard quantum mechanics. We
may speculate, that certain unaccounted for phenomena might be an
indication for, and might be explained by, such a special boundary
condition. For example, we may mention the inconsistencies in
measurements of cosmological parameters by different methods,
namely the dark matter puzzle.

\section{Summary}
We have suggested a \emph{Two-time interpretation} of quantum
mechanics. We have shown how a special final boundary condition
for the Universe, can effectively account for the observed wave
function reduction, while being consistent with standard quantum
mechanics formalism. The suggested interpretation is therefore one
which brings together quantum mechanics' empirical predictions and
formalism, thus solving the ``measurement problem''. What seems a
mystery to the single-time observer may come to a simple
resolution looking at our Universe from a two-time perspective.

\begin{acknowledgments}
We would like to thank Shmuel Nussinov, Benny Reznik, Daniel
Rohrlich and Lev Vaidman for reviewing this work. This research
was supported in part by grant 62/01-1 of the Israel Science
Foundation, established by the Israel Academy of Sciences and
Humanities, NSF grant PHY-9971005, and ONR grant N00014-00-0383.
\end{acknowledgments}

\end{document}